\documentclass[11pt]{article}
\usepackage{moriond,epsfig}

\def\Journal#1#2#3#4{{#1} {\bf #2}, #3 (#4)}

\def\be{\begin{equation}}
\def\ee{\end{equation}}
\def\bea{\begin{eqnarray}}
\def\eea{\end{eqnarray}}

\begin{document}
\vspace*{4cm}
\title{Testing an unifying view of Gamma Ray Burst afterglows}

\author{ M. Nardini$^1$, G. Ghisellini$^2$, G. Ghirlanda$^2$, A. Celotti$^1$ }

\address{$^1$ SISSA/ISAS, Via Beirut 2-4, 34014 Trieste, Italy
\\
$^2$  Osservatorio Astronomico di Brera, via E. Bianchi 46, IÐ23807 Merate, Italy  }

\maketitle\abstracts{
Four years after the launch the {\it Swift} satellite the nature of the
Gamma Ray Bursts (GRBs) broadband afterglow behaviour is still an open
issue ad the standard external shock fireball models cannot easily
explain the puzzling combined temporal and spectral optical to X--ray  behaviour
of a large number of afterglows. 
We analysed the rest frame de--absorbed and K--corrected optical and
X--ray multi--wavelength light--curves of a sample of 33 GRBs with known
redshift and optical extinction at the host frame. We modelled their
broadband behaviour as the sum of the standard forward shock emission
due to the interaction of a fireball with the circum--burst medium and
an additional component. We are able to obtain a good agreement with
the observed light--curves despite their complexity and diversity and
can also account for the lack of achromatic late times jet breaks in
several GRBs and explain the presence of chromatic breaks. Even if the
second component is treated in a phenomenological way, we
can identify it as a ``late prompt" emission due to a prolonged
activity of the central engine produced by a mechanism similar to the
one responsible for the early prompt emission. 
Our attempt can be considered as a first step towards the construction of
a  more physical scenario. A first important hint is that the
``late prompt" temporal decay is
intriguingly consistent with what expected with the the accretion
rate of fallback material. 
In order to test our model also from the spectral point of
view, we analysed the X--ray time resolved spectra and when possible the
evolution of the optical to X--ray spectral energy distribution.
All the events are found to be fully consistent with what predicted
by our model. Furthermore our analysis can give an alternative view to the
connection between the host galaxy dust reddening and the estimate of
the $N_{\rm H}$ column derived from the X--ray spectra.
}

\section{Introduction}

The launch of the {\it Swift} satellite represented a great 
improvement for the early time observations of Gamma Ray Bursts 
(GRBs) afterglows. 
The precise GRB localisation and the fast X--ray follow--up 
opened a new window on the understanding of the GRB afterglow emission. 
The behaviour of the X--ray light curves of a large fraction of events 
in the first thousands of seconds appeared much more complex than what 
had been observed and predicted in the pre--{\it Swift} era when it was 
possible to observe the X--ray afterglow only after some hours after 
the trigger.  
Most of the observed GRBs show a steep decay 
phase after the end of the prompt $\gamma$--ray emission that lasts for 
several dozens of seconds and is usually interpreted as the high 
latitude emission of the fireball
(Nousek et al.~\cite{no06}, Zhang et al.~\cite{zh06}). 
This sudden flux decay is 
followed by a phase in which the flux remains almost constant for 
a time that lasts from hundreds to hundred of thousand seconds 
depending on the specific GRB. 
This ``flat" phase triggered the interest 
of many groups and a large number of possible 
explanations have been proposed in literature. 
In Ghisellini 
et al.~\cite{GG09} a brief summary of some of the proposed models is given. 
After the end of the shallow phase (at a time called $T_{\rm A}$) 
the X--ray light curve changes behaviour and starts to decline as 
a power law $F\propto t^{-\alpha}$ with an index $\alpha\approx 1.3$ 
that represents the typical afterglow behaviour observed in the 
pre--{\it Swift} era.  
The optical light--curves instead seem 
not to trace the behaviour observed in the X--rays in a large number of events.

In this work we focus in particular on the model proposed by 
Ghisellini et al.~\cite{GG07}. In this model, after the standard 
prompt emission,  a prolonged activity of the central engine keeps 
producing shells with decreasing power and decreasing 
bulk Lorentz factor $\Gamma$. 
In this scenario during the shallow phase the decreasing Lorentz factor 
allows to see an increased portion of the emitting area. 
This effects ends 
at a characteristic time $T_{\rm A}$ when $1/\Gamma$ becomes 
equal to the jet opening angle $\theta_{\rm j}$.  
The observed radiation (both in the X--ray and in the optical bands)
during the shallow phase is thus explained as the superposition of 
a ``standard" forward shock afterglow emission and 
this second ``late prompt" component.

\section{Light--curve modelling}

\subsection{The sample}

We analysed a sample of long GRBs with known redshift, optical and 
XRT follow up, and a published estimate of the host galaxy dust 
absorption $A_{\rm V}^{\rm host}$. 
As at the end of March 2008 we 
found 33 GRBs fulfilling all our selection criteria. 
When possible, if multiple  $A_{\rm V}^{\rm host}$ estimates for an 
individual burst are present in literature, we choose  the one obtained 
analysing the optical data only, without assuming any connection 
with the X--rays data.  
We collected all the multi--band photometric 
data reported in literature for the GRBs in our sample and converted the 
observed magnitudes to monochromatic 
luminosities (de--reddened an K--corrected; see the relevant data and references 
in Ghisellini et al.~\cite{GG09}). 
The observed XRT 0.3--10 keV fluxes have been corrected for the 
Galactic and host frame $N_{\rm H}$ absorption and converted into rest frame 
K--corrected 0.3--10 keV luminosities.

\subsection{Phenomenological model}

 We modelled the rest frame luminosity light--curves as the sum of 
two separate components. 
The first one is modelled as a ``standard" 
forward shock afterglow component following the analytical description 
given in Panaitescu and Kumar~\cite{pk00}. 
This parametrisation needs 6 free parameters. 
Since we do not have a complete physical 
description of the second component we treated it in a completely 
phenomenological way with the aim of minimising the number of free 
parameters and to make a first step towards a more physical modelling. 
The second component spectral energy distribution is modelled as a 
smoothly joining double power--law whose shape, for simplicity,
is assumed not to evolve in time.
\begin{eqnarray}
L_{2^{nd}}(\nu, t) \, &=&\, L_0(t) \, \nu^{-\beta_{\rm x}}; \quad\quad
\qquad \nu>\nu_b \nonumber \\ 
L_{2^{nd}}(\nu, t) \, &=&\, L_0(t) \, 
\nu_{\rm b}^{\beta_{\rm o}-\beta_{\rm x}}\nu^{-\beta_{\rm o}}; \quad
\nu\le\nu_{\rm b}, 
\end{eqnarray}
where $L_0$ is a normalisation constant.  
The temporal behaviour of the second component is also described by a 
double power--law, with a break at $T_{\rm A}$ and with decay indices
$\alpha_{\rm flat}$ and 
$\alpha_{\rm steep}$ (before and after $T_{\rm A}$).
This modelling has 7 free parameters. 
Some of them can be well constrained directly by the observations 
(such as $T_{\rm A}$ and the spectral 
indices when the second component dominates the observed flux).

There is instead some degeneracy between the values of $\beta_{\rm o}$ 
and $\nu_{\rm b}$.       
In our modelling we did not take into account X--ray flares
and possible optical re--brightenings and bumps.

%-------------------------------------------
\begin{figure}[h!]
\begin{center}
\psfig{figure=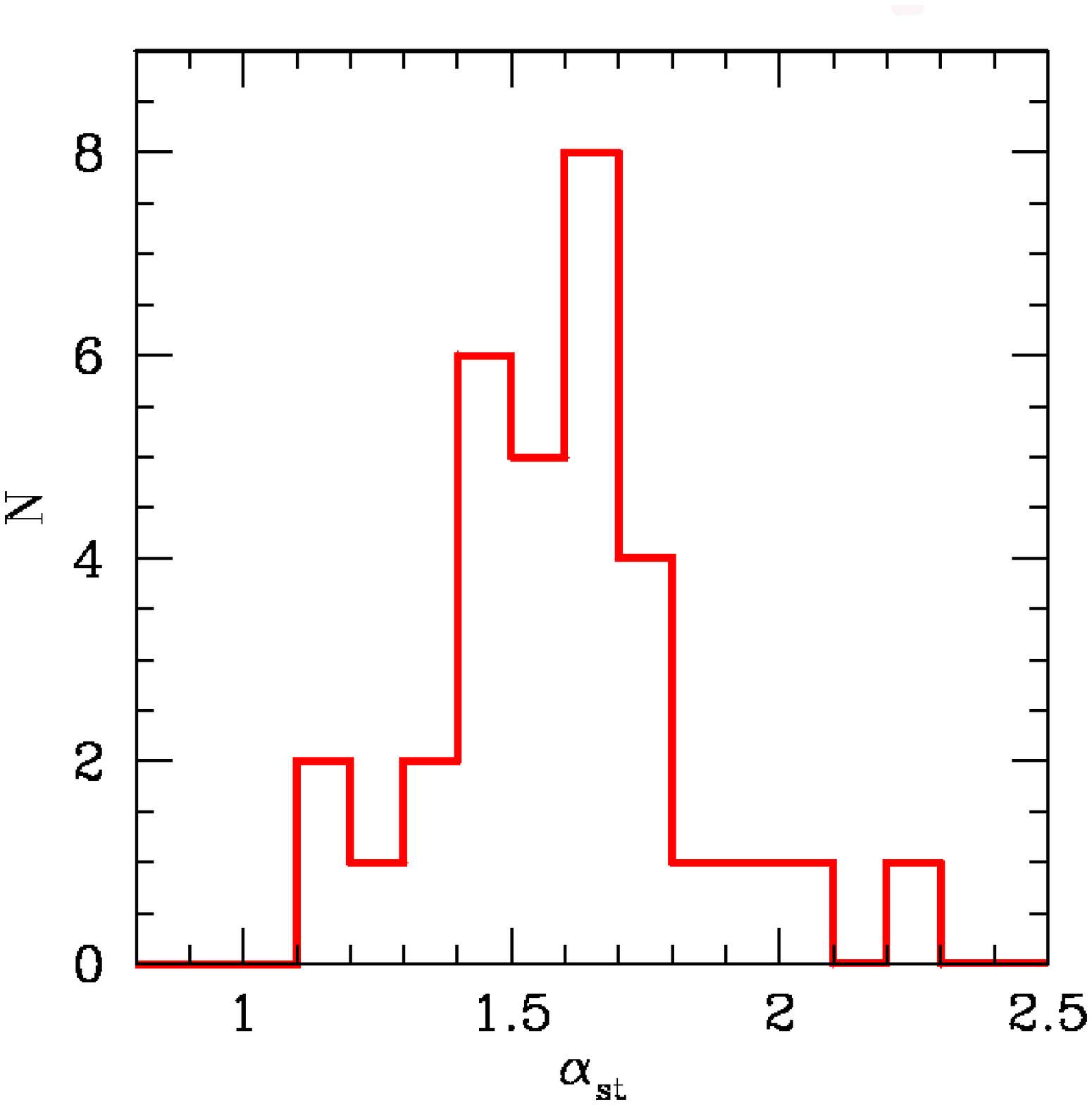,height=49mm}
\psfig{figure=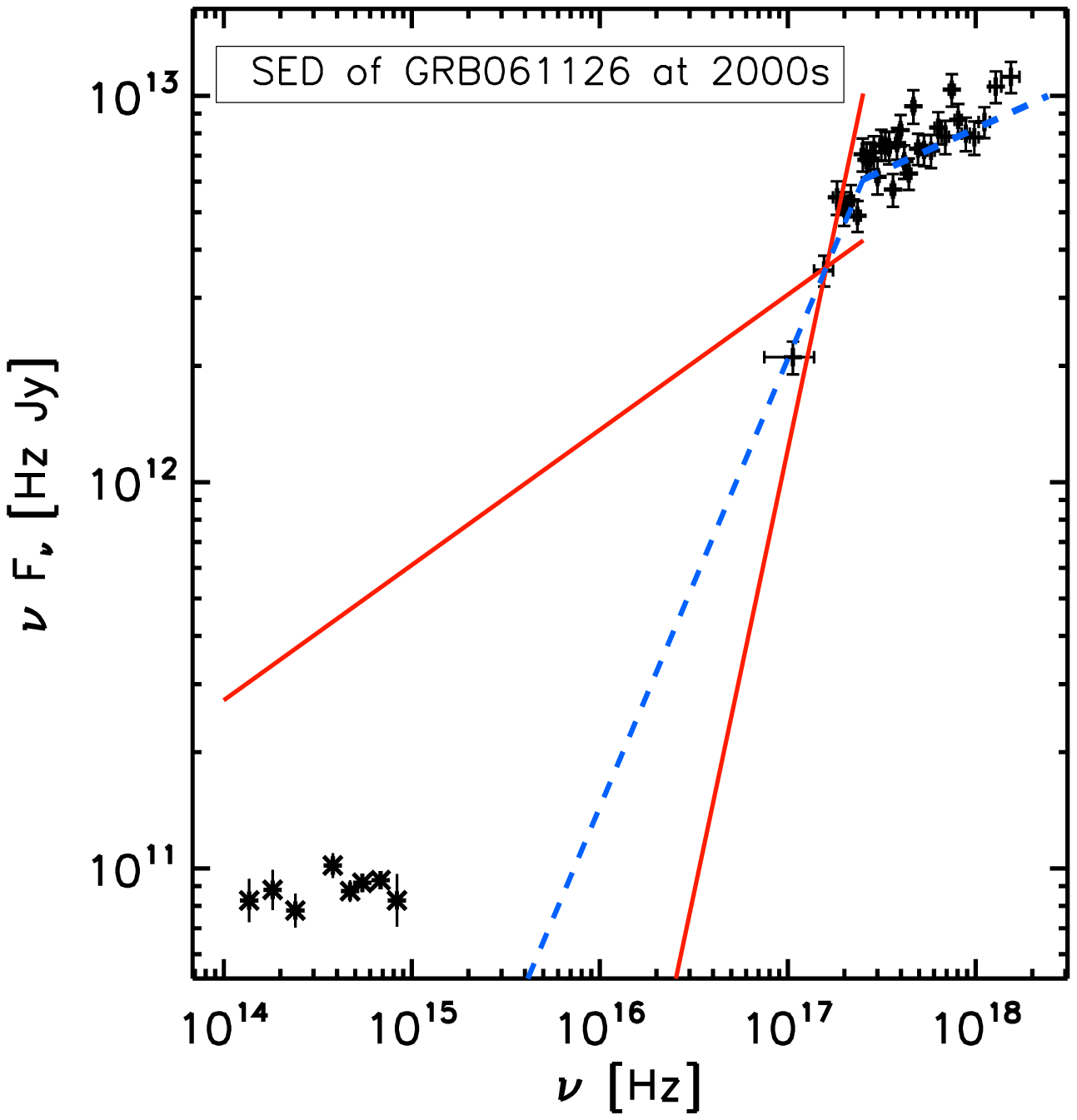,height=49mm}
\psfig{figure=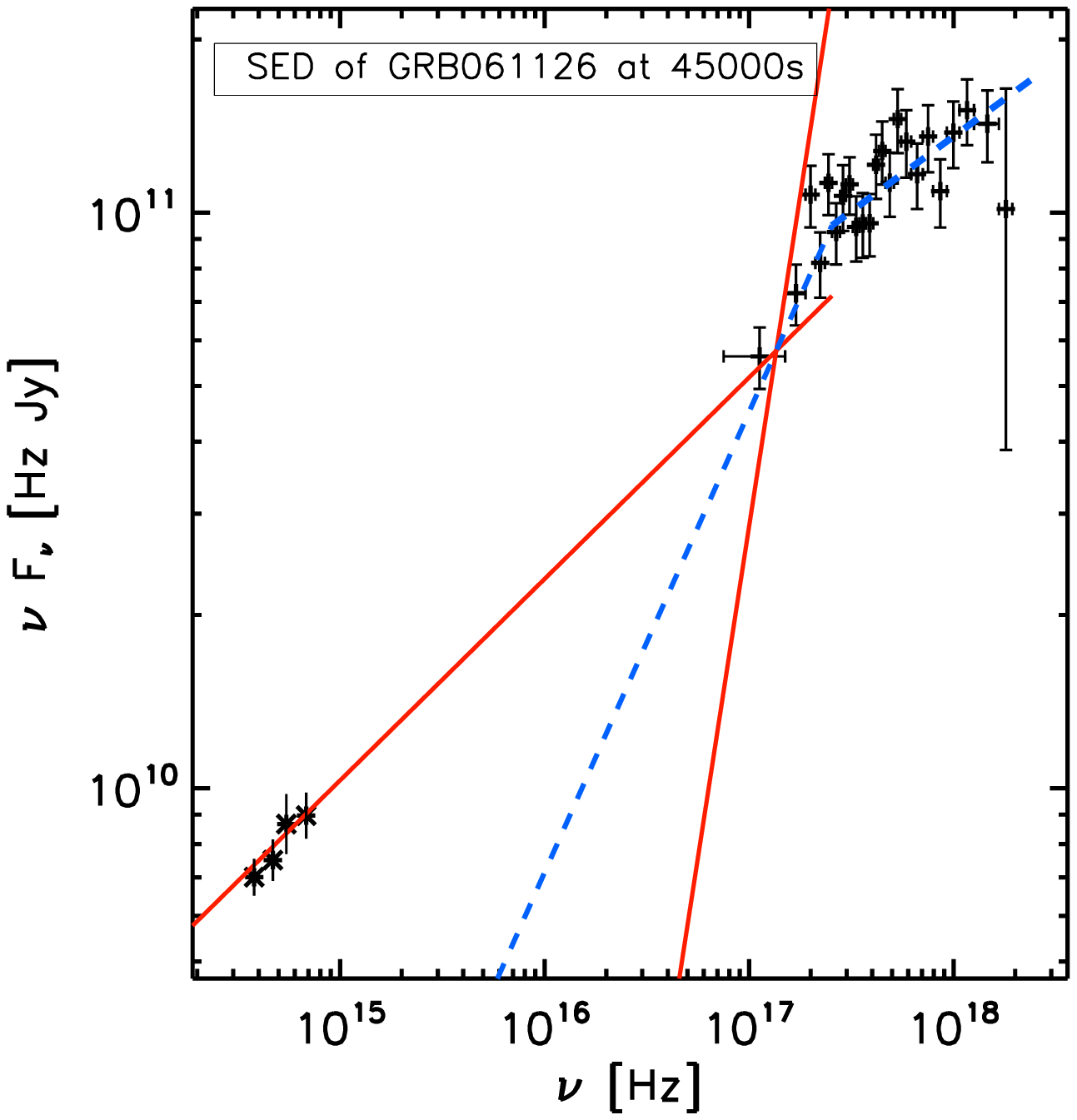 ,height=49mm}
\end{center}
\caption{
{\it Left panel}: The distribution of the decay index $\alpha_{\rm steep}$
of the second component. This is the decay index after $T_{\rm A}$.
{\it Central panel}: Early time SED of GRB 061126. 
{\it Right panel}: Late time SED of GRB 061126. 
}
\label{fig1}
\end{figure}   
% -------------------------------------------

\subsection{Results}

All the optical and X--ray light--curves of the GRBs in our sample
can be simultaneously reproduced rather well by our modelling. 
In 2 cases both the 
optical and X--ray light--curves are dominated by the second component while 
in 4 cases they are both dominated by the standard afterglow. 
The second component is dominant especially in the X--ray band (15 GRBs) 
while in the optical it dominates only in 3 GRBs. 
The afterglow dominates mainly in the optical (19 GRBs) while it 
is less important in the X--rays (6 GRBs). 
The remaining light--curves can be well described by 
a combination of the two components having almost the same 
importance or that dominate the light--curves in different time intervals. 
The distributions of the afterglow component parameters are similar to 
the ones obtained by Panaitescu and Kumar~\cite{pk02}. 
The distributions of the second component parameters show some interesting features. 
In particular the values of the post break  second component decay 
index $\alpha_{\rm steep}$ cluster around 1.6: this is remarkably 
close to $5/3$ (see fig. \ref{fig1}) that is the predicted decay of 
the accretion rate of fallback material onto the black hole 
(Chevalier~\cite{c89}).
It is also the average decay of the X--ray flare luminosity 
(Lazzati et al.~\cite{la08}). 
We also found an interesting correlation between the total energy emitted 
in $\gamma$--rays during the prompt event $E_{\gamma, \rm iso}$ and the 
energetics of the second component, estimated as $T_{\rm A}L_{\rm T_{\rm A}}$. 
This correlation is stronger than the one between $E_{\gamma, \rm iso}$ 
and the kinetic fireball energy $E_0$, implying that it is not simply 
due to the common redshift dependency. 

The small number of simultaneous breaks in optical and X--rays 
light curves in the {\it Swift} era opened a hot discussion about the 
nature of the jet breaks. 
In our scenario a jet break is expected only 
when the standard afterglow dominates the observed emission. 
When instead the flux is mainly produced by the second component, 
no jet break should be visible.
This can explain the lack of breaks at late times, 
the presence of chromatic breaks 
(when X--rays and optical bands are due to different components), 
and a post--break light curve decaying in a shallower way
than predicted (due to the contribution of the second component).

\section{Spectral check}

If the optical and the X--ray bands are produced by different 
processes there must be a spectral break in the  
spectral energy distribution between these two bands. 
For the light--curves modelling we assumed -- for simplicity -- that such a 
break always falls right in--between the optical and the X--ray bands,
but sometimes this break could occur in the observed XRT spectra. 
We then re--analysed all the XRT spectra of the GRBs in our sample (see Nardini et al.~\cite{MN09})
selecting time intervals not affected by prompt or high latitude 
emission or flaring activity. 
We first fitted the data with an absorbed single 
power--law model with frozen Galactic absorption plus a host 
frame absorption that was left free to vary. 
Our results are consistent with the ones found in literature and with the 
ones obtained using the automatic spectral analysis tool developed 
by Evans et al.~\cite{ev09}. 
We confirm the absence of spectral evolution around
$T_{\rm A}$, as predicted by the late prompt model that explains   
$T_{\rm A}$  as a purely geometrical effect. 
With the single power--law fitting 
we also confirm the inconsistency between the small $A^{\rm host}_{\rm V}$
derived in the optical 
and the usually large $N_{\rm H}^{\rm host}$ 
derived by the X--ray analysis,
if one assumes a standard $A_{\rm V}/N_{\rm H}$ relation 
(see e.g. Stratta et al.~\cite{st04}; Schady et al.~\cite{sc07}). 
We selected a sample of events with 
higher statistics spectra and we tried to fit them using a broken 
power--law model with the same two absorption components used in the 
case of the single power law fitting.  
In this GRB sub--sample we 
found 7 cases in which the presence of a break in the XRT observed 
spectra gives a better fit ($\Delta \chi ^2>5.5$) than the single 
power--law. In 8 GRBs instead the broken power--law model is excluded 
(break energy outside the considered energy range). 

For the 7 GRBs requiring a break in the X--ray band we can test 
if this break is consistent with what observed in 
the optical band.
We can also check if this is consistent with what predicted by our
double component model.
We then constructed and analysed the optical to X--rays SEDs at different times. 
The results are encouraging, since
for all bursts where the optical and X--ray fluxes 
were predicted to be produced by the same component, 
the optical lies on the extrapolation of the low energy index of the
X--ray spectra.
Instead, when the optical and X--rays were predicted to be produced
by different components, the extrapolation of the low energy spectral 
index to the optical band underestimates the observed flux. 
Furthermore, in GRB 061126 we can clearly see the transition between 
two phases.
The entire X--rays light--curve is dominated by the  
second component, while the optical flux is dominated
by the standard afterglow emission at early times.
Reassuringly, at these early times the optical SED is inconsistent with the 
extrapolation of the XRT spectrum. 
After about 2500 s, instead, the optical to X--ray SED can be described 
by a single component, and at these times the optical light--curve
was indeed predicted to be dominated by our second component (see fig. \ref{fig1}). 
In the 7 GRBs fitted 
with a broken power--law model the derived $N_{\rm H}^{\rm host}$  
is smaller than the one obtained with a single power--law model,
and closer to the values expected from $A_{\rm V}^{\rm host}$. 
On the other hand we derive very large $N_{\rm H}^{\rm host}$ columns 
also in some of the GRBs in which a broken power--law fitting is excluded. 
This means that the large  $N_{\rm H}/A_{\rm V}$ ratios observed in 
several GRBs can be sometimes due to an intrinsic spectral feature, 
but this can not be considered as a general solution of the large 
$N_{\rm H}/A_{\rm V}$ issue.


\section*{References} 
\begin{thebibliography}{99}

\bibitem{no06} J.A. Nousek {\it et al}, \Journal{ApJ}{642}{389}{2006}.

\bibitem{zh06} B. Zhang  {\it et al}, \Journal{ApJ}{642}{354}{2006}.

\bibitem{GG09} G. Ghisellini {\it et al}, \Journal{ApJ}{658}{L75}{2007}

\bibitem{GG07} G. Ghisellini {\it et al}, \Journal{MNRAS}{393}{253}{2009}

\bibitem{MN09} M. Nardini {\it et al}, MNRAS submitted (pre-print arXiv:0907.4157v1) 

\bibitem{pk00} A. Panaitescu \& P. Kumar, \Journal{ApJ}{543}{66}{2000}

\bibitem{pk02} A. Panaitescu \& P. Kumar, \Journal{ApJ}{571}{779}{2002}

\bibitem{c89} R.A. Chevalier, \Journal{ApJ}{346}{847}{1989}

\bibitem{la08} D. Lazzati, \Journal{MNRAS}{388}{L15}{2008}

\bibitem{ev09} P.A. Evans, accepted for publication in MNRAS  (pre-print arXiv:0812.3662v1) 

\bibitem{st04} G. Stratta, \Journal{ApJ}{608}{846}{2004}

\bibitem{sc07} P. Schady, \Journal{MNRAS}{377}{284}{2007}


\end{thebibliography}
\end{document}